\documentclass[conference]{IEEEtran}
\IEEEoverridecommandlockouts
\usepackage{cite}
\usepackage{amsmath,amssymb,amsfonts}
\usepackage{algorithmic}
\usepackage{graphicx}
\usepackage{textcomp}
\usepackage{xcolor}
\usepackage{url}
\def\BibTeX{{\rm B\kern-.05em{\sc i\kern-.025em b}\kern-.08em
    T\kern-.1667em\lower.7ex\hbox{E}\kern-.125emX}}
\begin{document}
\title{Spot-on: A Checkpointing Framework for Fault-Tolerant Long-running Workloads on Cloud Spot Instances}
\author{\IEEEauthorblockN{Ashley Tung$^\dagger$, Haiyan Wang$^\dagger$, Yue Li$^\dagger$, Zhong Wang$^*$, and Jingchao Sun$^\dagger$}
\IEEEauthorblockA{\textit{$^\dagger$MemVerge Inc., Milpitas, CA} \\
\textit{$^*$Department of Energy Joint Genome Institute, Berkeley, CA}\\
yue.li@memverge.com}
}

\maketitle
\begin{abstract}
Spot instances offer a cost-effective solution for applications running in the cloud computing environment. However, it is challenging to run long-running jobs on spot instances because they are subject to unpredictable evictions. Here, we present Spot-on, a generic software framework that supports fault-tolerant long-running workloads on spot instances through checkpoint and restart. Spot-on leverages existing checkpointing packages and is compatible with the major cloud vendors. Using a genomics application as a test case, we demonstrated that Spot-on supports both application-specific and transparent checkpointing methods. Compared to running applications using on-demand instances, it allows the completion of these workloads for a significant reduction in computing costs. Compared to running applications using application-specific checkpoint mechanisms, transparent checkpoint-protected applications reduce runtime by up to $40$\%, leading to further cost savings of up to $86$\%. 
\end{abstract}

%\begin{IEEEkeywords}
%checkpointing, Memory Machine, metagenome assembly, spot instances, cloud computing
%\end{IEEEkeywords}

\section{Introduction}
Major cloud vendors offer ``spot virtual machine (VM) instances'' that utilize spare computing resources at steep discounts~\cite{azure}\cite{aws}\cite{gcp}. However, a spot instance can be reclaimed during a resource shortage with a short notice seconds or minutes before a reclamation. Upon reclamation, all workloads running on the instances are terminated, and the instance is destroyed. This unpredictable nature makes it challenging to run long-running workloads on spot instances without checking points. This is not unlike Amazon EC2’s spot market used in Proteus(\cite{proteus}) and Tributary (\cite{tributary}). What sets Azure spot instances apart is that there is no need to bid for any new resources. Rather, the user is able to choose a VM size and simply have the option to turn it into a Spot instance.

Checkpoint solutions developed in high-performance computing systems can be adapted for the cloud environment~\cite{dmtcp}\cite{nersc}\cite{blobcr}. Both application-specific and transparent checkpointing technologies may be leveraged so that checkpoints can be made on one spot instance and moved to restart on another when the previous instance is reclaimed. However, to implement a practical solution that is user-friendly requires careful integration with all the cloud platforms and schedulers to properly schedule, store, transfer, and restart checkpoints. In this work, we implemented a practical framework called ``Spot-on'' by integrating with the major cloud vendor's spot instance scheduler to evaluate the impact of checkpointing mechanisms on running time and cost of long-running workloads. We used a case study, a long-running metagenome assembly workload (metaSPAdes, \cite{nurk2017metaspades}), to compare the checkpointing methods on Azure on-demand and spot instances. We found that both checkpointing methods enable fault-tolerance metaSPAdes workloads on spot instances to reduce cost. Compared to using application-specific checkpointing mechanisms on spot instances, metaSPAdes protected by transparent checkpointing takes less time to finish, which leads to further cost reductions. 

\section{Architecture and Design}
\begin{figure}[htbp]
\centerline{\includegraphics[width=\columnwidth]{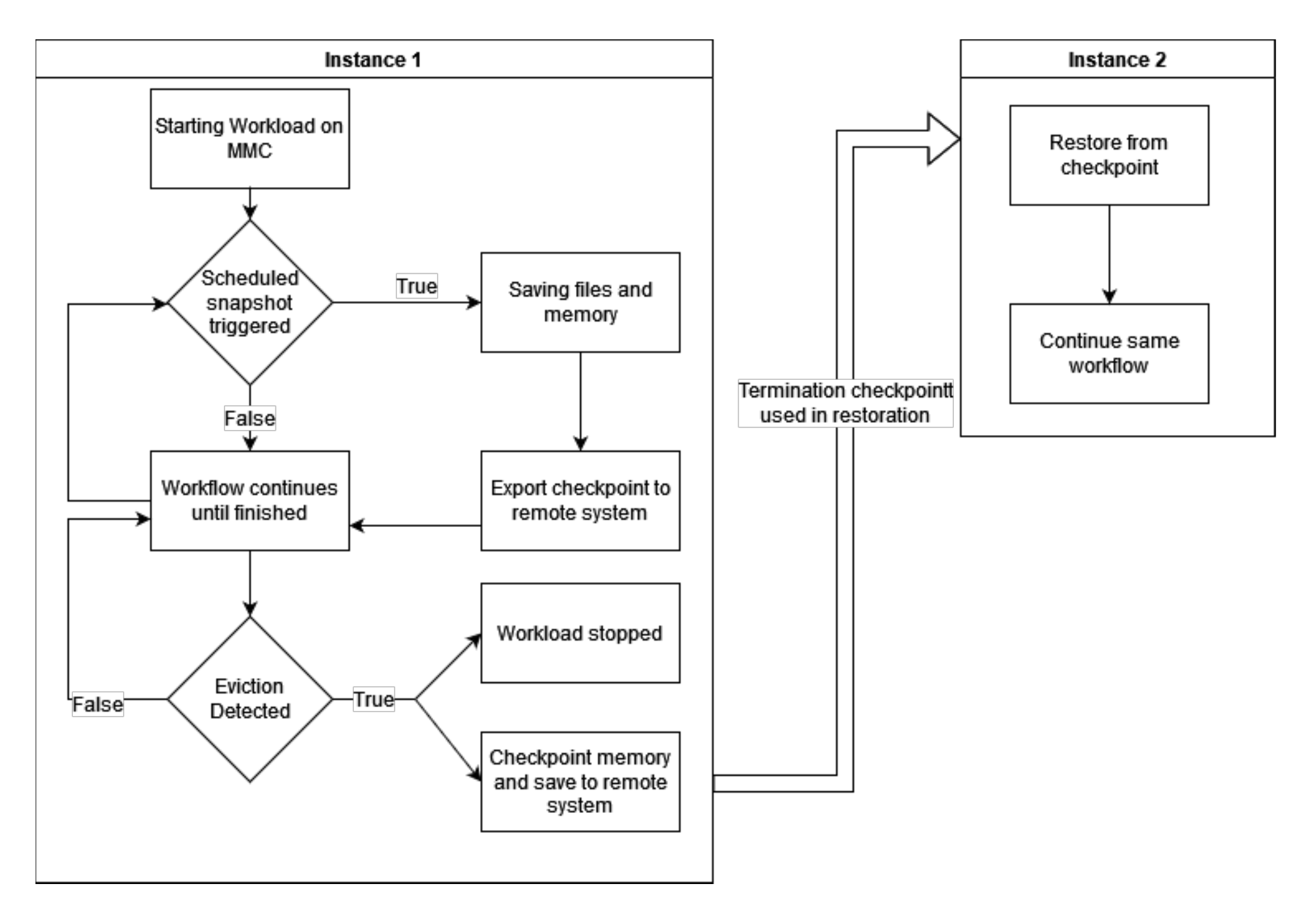}}
\caption{The Spot-on Checkpoint and Restart Workflow across spot instances.}
\label{fig:flow}
\end{figure}

The Spot-on checkpoint and restart workflow framework is illustrated in Fig.~\ref{fig:flow}.
When a workload is launched on the spot instance, a checkpoint coordinator, Spot-On, is launched simultaneously. Running the coordinator does not provide additional monetary cost to the user,as it is essentially a script running in parallel to metaSPAdes. The coordinator has the responsibility for checkpointing and restoration: it schedules periodic checkpointing and monitors VM eviction events using APIs provided by the cloud. Upon detecting an eviction event, the coordinator creates a ``termination checkpoint'' in addition to periodic checkpoints. Unlike the periodic checkpoints, termination checkpoints are opportunistic due to their possible failures caused by the short eviction notification (e.g. seconds to a few minutes). In this framework, both application-specific and transparent checkpointing are supported, and the coordinator is able to invoke the corresponding interfaces through its configuration files. 

After a spot instance is terminated, a new one is created manually or automatically through a cloud vendor's spot scheduling system or a separate job/resource scheduler (e.g., Slurm and LSF). The checkpoint coordinator then automatically searches for the most recent valid checkpoint and resumes the workload. The checkpoints taken from the terminated instance are transferred or shared with the new one through shared cloud storage services such as elastic block stores, network or distributed file systems, object, and blob stores.

\section{A Case Study in Metagenome Assembly}
metaSPAdes is one of the most widely used metagenome assemblers. As with many other metagenome assemblers, metagenome assembly with metaSPAdes is time-consuming, taking days or even weeks on large datasets. In this study, we used a public metagenomics dataset derived from samples taken at a wastewater treatment plant in Idaho~\cite{stalder2019linking}. This dataset has a total uncompressed size of $164.8$ GiB. We selected the first $50$ million reads with a size of around $4$ GiB to use in this study. metaSPAdes version 3.15.3~\cite{spadesgithub} was used with default settings and five k-mer sizes: $33,55,77,99$, and $127$.

We deployed workloads on Azure spot instances created by the virtual machine ``Scale Sets'' service\cite{scalesets}. Scale sets act as a VM pool manager that is capable of restarting new spot instances upon eviction of existing spot instances. For all tests, we used the same spot instance specifications: D8s v3 instance with 8 cores and $32$ GiB memory, CentOS 7.9. The spot price for the specified instance is $\$0.076$/hr, and its on-demand price is $\$0.38$/hr. For the scale sets' Custom Data, which runs scripts when a new instance is created, we use the coordinator to execute the workflow described in Section II (Fig.~\ref{fig:flow}).

\subsection{Checkpointing Methods}
To checkpoint metaSPAdes workloads, we used both its own checkpointing method and CRIU~\cite{criu}, which is a transparent checkpointing package. 
%metaSPAdes' checkpoints are made after major computation stages, such as error correction, K-mer calculation, or mismatch correction. For our workload and dataset, the error and mismatch correction is off by default. 
Compared to transparent checkpointing, application-specific checkpointing cannot be taken on demand. Checkpoints are shared between spot instances using Azure's NFS service, which charge \$$16.00$ per $100$GiB provisioned (\cite{file}).

\subsection{Triggering and Detecting Spot Evictions}
To monitor spot eviction notification, the checkpoint coordinator integrates Azure's REST API to access the ``Scheduled Events"~\cite{scheduledevents}, an Azure metadata service that gives the VM time to prepare for events, such as a spot instance reclaim/eviction. When a \texttt{GET} call is run within the VM to a specified endpoint, a JSON object is returned, listing the number of events and event types scheduled for the instance. Information is available via a non-routable IP so that it is not exposed outside the VM. An eviction notification is of type ``Preempt" and gives the VM a minimum of $30$ seconds to prepare for the eviction.

Since spot instance eviction is an unpredictable event, it is difficult to complete the evaluation based on true eviction events. Therefore, we use a command \texttt{simulate-eviction} provided by Azure CLI to artificially trigger a spot eviction. As the command produces the same event type as an actual Azure eviction, this command suffices to provide an Azure eviction.

\subsection{Preliminary Results}
We first evaluate the effectiveness of our framework in protecting the workload of metaSPAdes on spot instances. There is little difference in total execution time between running metaSPAdes with or without Spot-on when no checkpoint protection and no eviction were configured (Table.~\ref{tab1}), suggesting that Spot-on introduces little overhead. When we configured eviction time intervals at $60$ minutes or $90$ minutes, both application-native checkpoints and  transparent checkpoints offered protection and workloads were completed successfully. We set our transparent checkpointing times to every $10$ or $15$ minutes.
%
\iffalse
\begin{table}
\caption{\label{tab2}Evaluation results for metaSPAdes.\vspace{-0.6cm}}
\begin{center}
%
\resizebox{\columnwidth}{!}{\begin{tabular}{|l|l|l|l|l|l|l|l|l|l|l|}
\hline
K33 & K55 & K77 & K99 & K127 & Total & Spot Cost & On-demand Cost & Eviction & Checkpoint Type& Spot-on\\
\hline
33:50 & 38:53 & 39:51 & 40:19 & 30:33 & 3:03:26 & \$0.23 & \$1.16 & N/A & N/A & OFF\\
\hline
33:57&	39:03&	41:35&	40:41&	31:01&	3:05:32&	\$0.24& 	\$1.18 & N/A & N/A &ON\\
\hline
33:33 &	40:15&	57:16&	38:56&	46:14&	3:36:14&	\$0.28 	& \$1.33 & 	Every 90 min & Application&ON\\
\hline
29:22&	1:05:25&	1:03:03&	59:25&	51:07&	4:28:22&	\$0.34 	&\$1.70& Every 60 min & Application&ON\\
\hline
32:52&	37:03&	41:15&	39:53&	28:32&	2:59:35&	\$0.23& 	\$1.14&	Every 90 min & Transparent 30 min&ON\\
\hline
32:45&	38:13&	41:58&	39:50&	32:22&	3:05:08&	\$0.24& \$1.14&	 	Every 90 min & Transparent 15 min&ON\\
\hline
32:40&	38:52&	41:10&	39:45&	28:34&	3:01:01&	\$0.23 &	\$1.15&	Every 60 min & Transparent 30 min&ON\\
\hline
31:10&	38:15&	42:05&	40:01&	30:29	&3:02:00&	\$0.23& \$1.15& 	 	Every 60 min & Transparent 15 min&ON\\
\hline
\end{tabular}}
\label{tab1}
\end{center}
\vspace{-0.7cm}
\end{table}
\fi

\begin{table}
\caption{\label{tab2} Comparisons on execution time of metaSPAdes.\vspace{-0.6cm}}
\begin{center}
\resizebox{\columnwidth}{!}{\begin{tabular}{|l|l|l|l|l|l|l|l|l|}
\hline
K33 & K55 & K77 & K99 & K127 & Total & Eviction & Checkpoint Type& Spot-on\\
\hline
33:50 & 38:53 & 39:51 & 40:19 & 30:33 & 3:03:26 & N/A & N/A & OFF\\
\hline
33:57&	39:03&	41:35&	40:41&	31:01&	3:05:32 & N/A & N/A &ON\\
\hline
33:33 &	40:15&	57:16&	38:56&	46:14&	3:36:14 & 	Every 90 min & Application&ON\\
\hline
29:22&	1:05:25&	1:03:03&	59:25&	51:07&	4:28:22 & Every 60 min & Application&ON\\
\hline
32:52&	37:03&	41:15&	39:53&	28:32&	2:59:35 & Every 90 min & Transparent 30 min&ON\\
\hline
32:45&	38:13&	41:58&	39:50&	32:22&	3:05:08 & Every 90 min & Transparent 15 min&ON\\
\hline
32:40&	38:52&	41:10&	39:45&	28:34&	3:01:01 & Every 60 min & Transparent 30 min&ON\\
\hline
31:10&	38:15&	42:05&	40:01&	30:29	&3:02:00 & 	Every 60 min & Transparent 15 min&ON\\
\hline
\end{tabular}}
\vspace{-0.5cm}
\label{tab1}
\end{center}
\end{table}

\begin{figure}[htbp]
\centerline{\includegraphics[width=\columnwidth]{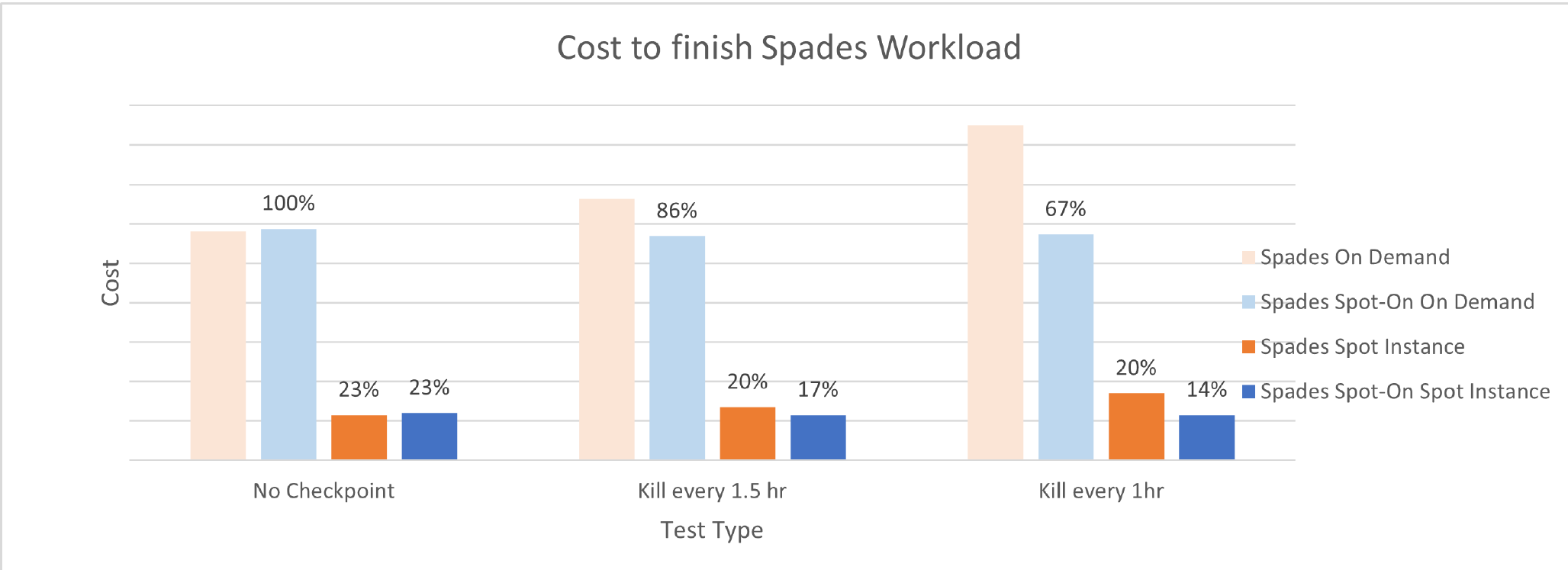}}
\caption{Cost comparisons when running using on-demand versus spot instances.}
\label{fig2}
\end{figure}

\begin{figure}[htbp]
\centerline{\includegraphics[width=\columnwidth]{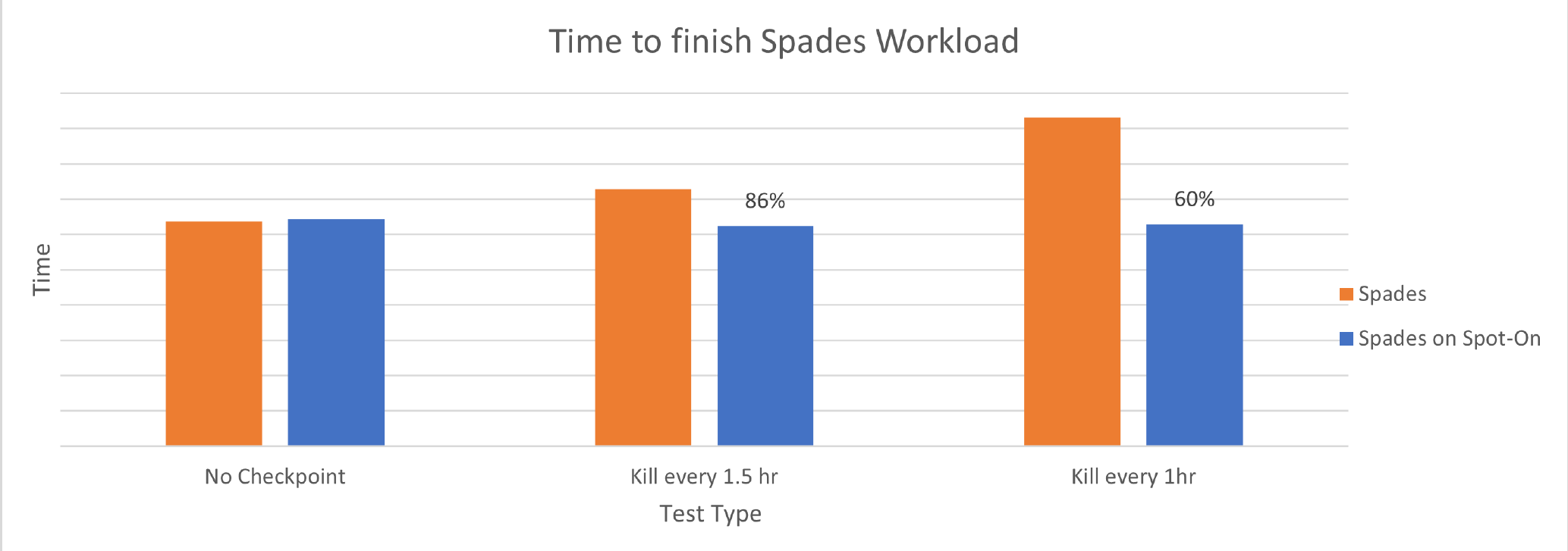}}
\caption{Execution time comparisons when running with application-native checkpointing and transparent checkpointing on spot instances.}
\label{fig3}
\end{figure}

Assuming that the same VM is used (i.e. using the same number of cores and memory), VM size and Azure NFS service, running metaSPAdes on checkpoint-protected spot instances saves $77$\% of costs over on-demand instances, simply from the price cuts between on-demand and spot instances alone (Fig.~\ref{fig2}). Additionally, running metaSPAdes with transparent checkpointing on these spot instances can save up to $86$\% of costs over the same workload on on-demand instances without checkpointing. Transparent checkpointing also adds about additional $15$-$40$\% time savings over application checkpoint (Fig~\ref{fig3}). Naturally, had eviction time interval been shorter, the percentage of time and cost saved by running metaSPAdes with Spot-On transparent checkpointing on Spot Instances would increase further compared to running metaSPAdes baseline using On-demand VMs.

\section{Conclusion}
 We provided a flexible framework, Spot-on, that integrates with the cloud computing environment and supports both application-specific and transparent checkpointing methods. A detailed case study in metagenome assembly with this framework suggests that leveraging checkpointing not only provides fault tolerance, but also reduces computing cost, especially when using transparent checkpointing on spot instances. Besides preventing unpredictable eviction events of the spot instances with reduced cost, the transparent checkpointing Memory Machine Checkpointing can potentially have other advantages. It can support other types of interruption, such as out-of-memory, in which case the workload can be resumed on a larger instance from a checkpoint. Some long-running jobs relying solely on application-specific checkpointing may never be able to complete if the time between application checkpointing is longer than the lifetime of a spot instance. The transparent checkpointing can effectively overcome this limit. 

\bibliography{reference}

\end{document}